\documentclass[12pt, letterpaper, twoside]{article}
\usepackage[utf8]{inputenc}
\usepackage{graphicx}

\title{Complexity in Mathematics Education}
\author{Brent Davis \\
Werklund School of Education \\
University of Calgary \\
Calgary, Alberta, Canada \\
Email: brent.davis@ucalgary.ca\\ \\
Pratim Sengupta
\\
Werklund School of Education
\\
University of Calgary
\\
Calgary, Alberta, Canada
\\
Email: pratim.sengupta@ucalgary.ca\\ \\}

\date{Author Note: This chapter has been accepted for publication in: \textbf{Lerman, S. (Ed.), Encyclopedia of Mathematics Education, Springer.}}
 
\begin{document}
 
\begin{titlepage}
\maketitle
\end{titlepage}

\section{Introduction}

Over the past half-century, ``complex systems'' perspectives have risen to prominence across many academic domains in the sciences, engineering and the humanities. Mathematics was among the originating domains of complexity research. Education has been a relative latecomer, and so perhaps not surprisingly, mathematics education researchers have been leading the way in the field.

There is no unified definition of complexity, principally because formulations emerge from the study of specific phenomena. One thus finds quite focused definitions in such fields as mathematics and software engineering, more indistinct meanings in chemistry and biology, and quite flexible interpretations in the social sciences (cf. Mitchell 2009). Because mathematics education reaches across several domains, conceptions of complexity within the field vary from the precise to the vague, depending on how and where the notion is taken up. Diverse interpretations do collect around a few key qualities, however. In particular, complex systems adapt and are thus distinguishable from complicated systems. A complicated system is one that comprises many interacting components, and whose global character can be adequately described and predicted by specifying the rules of operation of the individual parts. A complex system comprises many interacting agents, and \emph{emergence} of global behaviors that cannot be adequately predicted by simply specifying the rules of the individual agents is a central characteristic of such systems. Some popularly cited examples of complex, emergent phenomena include anthills, economies, and brains, which are more than the linear sum of behaviors of individual ants, consumers, and neurons. In brief, whereas the opposite of complicated is simple, opposites of complex include reducible and decomposable. Hence prominent efforts toward a coherent, unified description of complexity revolve around such terms as emergent, noncompressible, multi-level, self-organizing, context-sensitive, and adaptive.

This entry is organized around four categories of usage within mathematics education -- namely, complexity as: an epistemological discourse, an historical discourse, a disciplinary discourse, and a pragmatic discourse. 

\section{Complexity as an epistemological discourse}

Among educationists interested in complexity, there is frequent resonance with the notions that a complex system is one that knows (i.e., perceives, acts, engages, develops, etc.) and/or learns (adapts, evolves, maintains self-coherence, etc.). This interpretation reaches across many systems that are of interest among educators, including physiological, personal, social, institutional, epistemological, cultural, and ecological systems. Unfolding from and enfolding in one another, it is impossible to study one of these phenomena without studying all the others.

This is a sensibility that has been well represented in the mathematics education research literature for decades in the form of varied theories of learning. Among others, radical constructivism, socio-cultural theories of learning, embodied, and critical theories share essential characteristics of complexity. That is, they all invoke bodily metaphors, systemic concerns, evolutionary dynamics, emergent possibilities, and self-maintaining properties. Of particular relevance is the recent emphasis on intersectionality as a key element of critical race and gender theories, which explicitly situates our experiences of knowing and learning in mathematics classrooms as emergent from our simultaneous positions of marginalization and privilege, as well as the interplay between historical, institutional and social forces and individual desires (Levya 2017).

As illustrated in Figure 1, when learning phenomena of interest to mathematics educators are understood as nested systems, a range of theories become necessary to grapple with the many issues the field must address. A pedagogy for knowing and doing mathematics that is epistemologically committed to complexity necessitates insights in the form of multi-level and diverse models of the complex dynamics of knowing and learning. More significantly, perhaps, by introducing the systemic transformation into discussions of individual knowing and collective knowledge, complexity not only enables but compels a consideration of the manners in which knowers and systems of knowledge are co-implicated (Davis and Simmt 2006).

\includegraphics[width=1\textwidth]{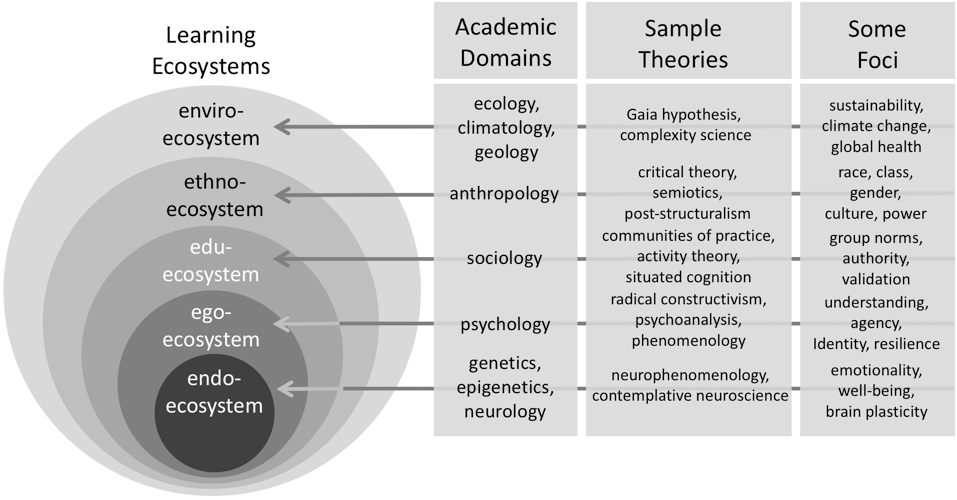}

Figure 1. Some of the nested complex systems of interest to mathematics educators

\section{Complexity as an historical discourse}

School mathematics curricula is commonly presented as a-historical and a-cultural. Contra this perception, complexity research offers an instance of emergent mathematics that has arisen and that is evolving in a readily perceptible time frame. As an example of what it describes -- a self-organizing, emergent coherence -- complexity offers a site to study and interrogate the nature of mathematics, interrupting assumptions of fixed and received knowledge.

To elaborate, the study of complexity in mathematics reaches back the late 19th century when Poincaré conjectured about the three-body problem in mechanics. Working qualitatively, from intuition Poincaré recognized the problem of thinking about complex systems with the assumptions and mathematics of linearity (Bell 1937). The computational power of mathematics was limited the calculus of the time; however, enabled by digital technologies of the second half of the 20th century, such problems became tractable and the investigation of dynamical systems began to flourish. With computers, experimental mathematics was born and the study of dynamical systems led to new areas in mathematics. Computational modeling made it possible to model and simulate the behaviour of a function over time by computing thousands and hundreds of thousands of iterations of the function. Numerical results were readily converted into graphical representations (the Lorenz attractor, Julia sets, bifurcation diagrams) which in turn inspired a new generation of mathematicians, scientists and human scientists to think differently about complex dynamical systems. Further advances in computing in the form of parallel and distributed computing and multi-agent modeling enabled scientists and mathematicians to simulate emergent phenomena by modeling simultaneous interactions between thousands of interacting agents (Mitchell 2009). Through such efforts, since the mid-20th century, as mathematicians, physical and computer scientists were exploring dynamical systems (e.g., Smale, Prigogine, Lorenz, Holland, etc.), their work and the work of biologists, engineers and social scientists became progressively more intertwined and interdisciplinary (Gilbert and Troitzsch 2005; McLeod and Nersessian 2016).

In brief, the emergence of complexity as a field of study foregrounds that mathematics might be productively viewed as a humanity. More provocatively, the emergence of a mathematics of implicatedness and entanglement alongside the rise of a more sophisticated understanding of humanity's relationship to the more-than-human world might be taken as an indication of the ecological character of mathematics knowledge.

\section {Complexity as a disciplinary discourse}

A common criticism of contemporary grade school mathematics curriculum is that little of its content is reflective of mathematics developed after the 16th or 17th centuries, when publicly funded and mandatory education spread across Europe. A deeper criticism is that the mathematics included in most pre-university curricula is fitted to a particular worldview of cause--effect and linear relationships. Both these concerns might be addressed by incorporating complexity-based content into programs of study.

Linear mathematics held sway at the time of the emergence of the modern school -- that is during the Scientific and Industrial Revolutions -- because it lent itself to calculations that could be done by hand. Put differently, linear mathematics was first championed and taught for pragmatic reasons, not because it was seen to offer accurate depictions of reality. Descartes, Newton and their contemporaries were well aware of nonlinear phenomena. However, because of the intractability of many nonlinear calculations, when they arose they were routinely replaced by linear approximations. As textbooks omitted nonlinear accounts, generations of students were exposed to over-simplified, linearized versions of natural phenomena. Ultimately that exposure contributed to a resilient worldview of a clockwork reality.

However, recent advances in computational modeling have made it possible for complex phenomena that are traditionally taught in post-secondary levels, to be easily accessible to much younger learners. With the ready access to similar technologies in most school classrooms within a culture of ubiquitous computation, there is now a growing call for deep, curricular integration of computer-based modeling and simulation in K--12 mathematics and science classrooms (Wilkerson and Wilensky 2015; Sengupta et al. 2015). Efforts for such integration fundamentally rely on learners iteratively designing, evaluating and re-designing mathematical models as the pedagogical approach, using agent-based modeling languages and platforms (e.g., Scratch, Agentsheets, NetLogo, ViMAP, etc.). In agent-based modeling, learners can simulate the relevant mathematical behaviors by programming the on-screen behavior of computational agents (e.g., the Logo turtle) using body-syntonic commands (e.g., move forward, turn, etc.). Emergence, in such computational models, is simulated as the aggregate-level outcome that arises from the interactions between many individual-level computational agents. The creator of the first such modeling language (Logo), Papert (1980) argued that agent-based modeling can create space in secondary and tertiary education for new themes such as recursive functions, fractal geometry and modeling of complex phenomena with mathematical tools such as difference equations, iterations, etc. Others (e.g., English 2006, Lesh and Doerr 2003) have advocated for similarly themed content, but in a less calculation-dependent format, arguing that the shift in sensibility from linearity to complexity is more important than the development of the computational competencies necessary for sophisticated modeling. In either case, the imperative is to provide learners with access to the tools of complexity, along with its affiliated domains of fractal geometry, chaos theory, and dynamic modeling.

New curriculum in mathematics is emerging. More profoundly, when, how, who and where we teach are also being impacted by the presence of complexity sensibilities in education because they are a means to nurture emergent possibility.

\section{Complexity as a pragmatic discourse}

To recap, complexity has emerged in education as a set of mathematical tools for analysing phenomena; as a theoretical frame for interpreting activity of adaptive and emergent systems; as a new sensibility for orienting oneself to the world; and for considering the conditions for emergent possibilities leading to more productive, ``intelligent'' classrooms. In the last of these roles, complexity might be regarded as the pragmatic discourse -- and of the applications of complexity discussed here, this one may have the most potential for affecting school mathematics by offering guidance for structuring learning contexts and re-shaping disciplinary pedagogies. Three key insights have emerged in the literature that can guide pragmatic action in the K--12 classroom. First, complexity offers direct advice for organizing classrooms to support the individual-and-collective generation of insight -- by, for example, nurturing the common experiences and other redundancies of learners while making space for specialist roles, varied interpretations, and other diversities. For example, participatory simulations, in which each learner can themselves play the role of an agent in complex system using embodied, physical and computational forms of modeling, have been shown to be effective pedagogical approaches for modeling emergent mathematical behaviors by highlighting and integrating both individual and collective insight (e.g., Colella 2000). Second, the emphasis on such participatory forms of mathematical modeling, in the context of modeling complex phenomena, can act as a bridge across disciplines (e.g., biology and mathematics education, see Dickes et al. 2016). A third key insight is the notion of reflexivity across disciplines -- that is, conceptual development within each scientific, engineering and mathematical discipline can be deepened further when relevant phenomena are represented as complex systems using mathematical modeling in ways that also highlight key practices of engineering design such as design thinking (Sengupta et al. 2013).

As complexity becomes more prominent in educational discourses and entrenched in the infrastructure of ``classrooms'', mathematics education can move from an individualistic culture to one of cooperation and collaboration, and from mono-disciplinarity towards inter- and trans-disciplinarity. These, in turn, have entailments for the outcomes of schooling as evident in movements from disciplinary ideas to crosscutting practices, from independent workers to team-based workplaces, and from individual knowing to social action.

\section{Cross References}
Design Research in Mathematics Education, Mathematical Modelling and Applications in Education, Technology and Curricula in Mathematics Education, Technology~Design in Mathematics Education, Theories of Learning Mathematics

\section{References}
\begin{itemize}

\item Bell ET (1937) Men of mathematics: The lives and achievements of the great mathematicians from Zeno to Poincaré. Simon and Schuster, New York.

\item Colella, V (2000) Participatory simulations: Building collaborative understanding through immersive dynamic modeling. The Journal of the Learning Sciences, 9(4): 471--500.

\item Davis B, Renert M (2013) The math teachers know: Profound understanding of emergent mathematics. Routledge, New York.

\item Davis B, Simmt E (2006) Mathematics-for-teaching: An ongoing investigation of the mathematics that teachers (need to) know. Educational Studies in Mathematics 61(3): 293--319.

\item Dickes, AC, Sengupta, P, Farris, AV, Basu, S (2016) Development of mechanistic reasoning and multilevel explanations of ecology in third grade using agent‐based models. Science Education,100(4): 734--776.

\item English L (2006) Mathematical modeling in the primary school: Children's construction of a consumer guide. Educational Studies in Mathematics, 62(3): 303--329.

\item Gilbert, N, Troitzsch, K (2005).~Simulation for the social scientist. McGraw-Hill Education, New York.

\item Lesh R, Doerr H (2003) (eds) Beyond constructivism: Models and modelling perspectives on mathematics problem solving learning and teaching. Lawrence Erlbaum Associates, Mahwah, NJ.

\item MacLeod, M, Nersessian, NJ (2016) Interdisciplinary problem-solving: Emerging modes in integrative systems biology.European journal for Philosophy of Science, 6(3): 401--418.

\item Mitchell M (2009) Complexity: a guided tour. Oxford University Press, Oxford, UK.

\item Mowat E, Davis B (2010) Interpreting embodied mathematics using network theory: Implications for mathematics education. Complicity: An International Journal of Complexity and Education, 7(1): 1--31.

\item Sengupta, P, Dickes, A, Farris, AV, Karan, A, Martin, D, Wright, M (2015) Programming in K--12 science classrooms.~Communications of the ACM, 58(11): 33--35.

\item Sengupta, P, Kinnebrew, JS, Basu, S, Biswas, G, Clark, D (2013) Integrating computational thinking with K--12 science education using agent-based computation: A theoretical framework.Education and Information Technologies, 18(2): 351--380.

\item Papert, S (1980) Mindstorms: Children, computers, and powerful ideas. Basic Books, New York.

\item Leyva, LA (2017) Unpacking the male superiority myth and masculinization of mathematics at the intersections: A review of research on gender in mathematics education. Journal for Research in Mathematics Education, 48(4): 397--433.

\item Wilkerson-Jerde, MH, Wilensky, UJ (2015) Patterns, probabilities, and people: Making sense of quantitative change in complex systems. Journal of the Learning Sciences,~24(2): 204--251.

\end{itemize}
 
\end{document}